\documentstyle[aps,preprint]{revtex}
\input{epsf}
\tighten
\begin{document}

\draft

\title{Quadrupole shape invariants in the interacting boson model} 

\author{V. Werner$\,^1$, N. Pietralla$\,^1$, P. von Brentano$\,^1$, 
        R.F. Casten$\,^2$, and R.V. Jolos$\,^{1,3}$
        }

\address{$^1\,$ Institut f\"ur Kernphysik, Universit\"at zu K\"oln, 
                50937 K\"oln, Germany}
\address{$^2\,$ Wright Nuclear Structure Laboratory, Yale University, 
	        New Haven, CT, USA}
\address{$^3\,$ Bogoliubov Laboratory, Joint Institute for Nuclear Research, 
                141980 Dubna, Russia}
 
\date{\today}

\maketitle

\begin{abstract}

In terms of the Interacting Boson Model, shape invariants for the ground state,
formed by quadrupole moments up to sixth order, are studied in the 
dynamical symmetry limits and over the whole structural range of the 
\hbox{IBM-1}. 
The results are related to the effective deformation parameters and their
fluctuations in the geometrical model. New signatures that can distinguish
vibrator and $\gamma$-soft rotor structures, and one that is related to 
shape coexistence, are identified.

\end{abstract}


\narrowtext

Nuclei are often regarded as drops of nuclear matter as in
the geometrical model of Bohr and Mottelson. Having a view of nuclei
as such geometrical objects leads directly to the importance of 
possible deformations of nuclei. The most important deformation of nuclei at
low energies is the quadrupole deformation to which we restrict our 
discussion. These quadrupole deformations are of special
interest as they enable us to make predictions of nuclear 
properties such as energies or $E2$ transition strengths of the lowest 
excited states.

Conversely one can deduce information about nuclear deformations
by observing $E2$ transition matrix elements. Indeed from a complete set of
$E2$ matrix elements one can calculate model independent moments and higher 
order moments of the quadrupole operator, tensorially coupled to a scalar -- 
the shape invariants. Shape invariants were first introduced by
Kumar \cite{Kum72} and Cline \cite{Cli86} in the discussion of a large set 
of $E2$ matrix elements
obtained in Coulomb excitation experiments. Calculating shape invariants 
in the geometrical model shows their connection to the deformation
parameters $\beta$ and $\gamma$ used by Bohr and Mottelson or, to be more
precise, to effective values $\beta_{\rm eff}$ and $\gamma_{\rm eff}$ and the 
fluctuations of those. Recently Jolos {\em et al.} \cite{Jol97,Pal98} have 
introduced approximation formulae to the lowest shape invariants in the 
framework of the newly developed $Q$-phonon scheme 
\cite{Ots94,Pie94,Pie95,Pie98}. These approximations now make it 
possible to determine approximate values of the shape invariants from data by 
using only a few absolute $B(E2)$ values.

This is a substantial result since the advent of radioactive beams opens up
entirely new nuclear regions for study but, at the same time, the very low
intensities of such beams means that data will be sparse and that nuclear
structure information must be obtained from fewer and simpler-to-obtain data.
Hence the importance of the approximations to the Q-invariants which allow
estimates not only of basic deformation parameters such as $\beta$ and
$\gamma$, but of higher moments related to the stiffness of the potential
in $\beta$ and $\gamma$ and to the amount of zero point motion. Such 
information has seldom if ever been available from any nuclear data. With
these approximation formulae, they are now accessible from simple data.

It is therefore important to develop a global view of how these shape
invariants behave as a function of structure so that they can be effectively 
used as signatures of structure. It is the purpose of this Rapid Communication
to map out for the first time the behaviour of the five essential invariants,
as well as several related quantities, over the full range of nuclear
structure. To do so we will use the algebraic Interacting Boson Model (IBM) 
\cite{AriIac75,IacAri87} to 
study the behaviour of shape invariants in and between the dynamical 
symmetry limits of the IBM.
Formulae will be given to transform the shape invariants into effective
deformation parameters $\beta$ and $\gamma$. The values derived from the 
algebraic model will be compared to values in the appropriate limiting cases 
of the geometrical model.

Shape invariants are formed by the isoscalar electric quadrupole
operator, which is also the $E2$ transition operator in the Consistent Q
Formalism (CQF) \cite{WarCas82},
\begin{equation}
\label{eq:TE2eqQ}
T(E2) = Q = e_B\ Q^{IBM} \ ,
\end{equation}
where $Q^{IBM}$ is the quadrupole operator in the IBM 
\begin{equation}
\label{eq:qchi}
Q^{IBM}=Q^{\chi}=s^+\tilde{d}+d^+ s +\chi [d^+\tilde{d}]^{(2)}
\end{equation}
and $e_B$ is the effective boson charge which is fixed for a given nucleus. 
We define moments up to sixth order of the quadrupole operator in the ground
state as
\begin{eqnarray}
\label{eq:defq2}
q_2 = & & \langle 0^+_1|(Q\cdot Q)|0^+_1\rangle \\
\label{eq:defq3}
q_3 = & \sqrt{\frac{35}{2}} \  & |\langle 0^+_1|[QQQ]^{(0)}|0^+_1\rangle| \\
\label{eq:defq4}
q_4 = & & \langle 0^+_1|(Q\cdot Q) \ (Q\cdot Q)|0^+_1\rangle \\
\label{eq:defq5}
q_5 = & \sqrt{\frac{35}{2}} \  & |\langle 0^+_1|(Q\cdot Q) \ [QQQ]^{(0)}|0^+_1\rangle| \\
\label{eq:defq6}
q_6 = & \frac{35}{2} & \langle 0^+_1|[QQQ]^{(0)} \ [QQQ]^{(0)}|0^+_1\rangle \ ,
\end{eqnarray}
where a dot denotes a scalar product and $[QQQ]^{(0)}$ abbreviates 
the tensor coupling
$[Q\times [Q\times Q]^{(2)}]^{(0)}$. We should note that $q_2$ is equal to 
the total absolute $E2$ excitation strength from the ground state 
\begin{equation}
q_2 = \sum_j B(E2;0^+_1 \rightarrow 2^+_j) \ .
\end{equation}
$q_2$ will be the only quantity in our discussion where an absolute value, 
namely the effective boson charge $e_B$, appears.

With the moments (\ref{eq:defq2}--\ref{eq:defq6}) we define the 
relative dimensionless shape invariants by normalizing to an appropriate
power of $q_2$
\begin{equation}
K_n = \frac{q_n}{{q_2}^{n/2}}  \ \mbox{for } \  n\in\{3,4,5,6\} \ .
\end{equation}
The quantities $K_n$ do not depend on the effective boson charge $e_B$.
The shape invariants $K_n$ differ from earlier definitions of
Jolos {\em et al.} \cite{Jol97} 
by normalization constants or tensor coupling. 
In the present definitions no
value may become infinite and all shape invariants are exactly equal to unity 
in the limit of the rigid symmetric rotor or, in terms of the IBM, the 
$SU(3)$ limit for any boson number $N$.

For the calculation of the shape invariants it is convenient to write
the expressions for the quadrupole moments (\ref{eq:defq2}--\ref{eq:defq6})
as sums over $E2$ matrix elements. Therefore the tensor properties of
the quadrupole operator are taken into account and the unity operator 
\hbox{${\bf 1} = \sum_{J,i,M}|J_i M\rangle \langle J_i M|$} is inserted 
between every pair of quadrupole operators. Using the Wigner-Eckert
theorem and the unitarity relation of Clebsch Gordan coefficients 
it is possible to write the moments $q_n$ as
\begin{eqnarray}
\label{eq:q2sum}
q_2 = & & \sum_i \langle 0^+_1 \parallel Q\parallel 2^+_i\rangle 
		 \langle 2^+_i \parallel Q\parallel 0^+_1 \rangle \\
\label{eq:q3sum}
q_3 = & \sqrt{\frac{7}{10}} \  & |\sum_{i,j} 
		  \langle 0^+_1 \parallel Q\parallel 2^+_i\rangle 
                  \langle 2^+_i \parallel Q\parallel 2^+_j\rangle 
 		  \langle 2^+_j \parallel Q\parallel 0^+_1 \rangle| \\
\label{eq:q4sum}
q_4 = & & \sum_{i,j,k} \langle 0^+_1 \parallel Q\parallel 2^+_i\rangle 
		       \langle 2^+_i \parallel Q\parallel 0^+_j \rangle 
		       \langle 0^+_j \parallel Q\parallel 2^+_k\rangle \nonumber \\
 & & 		       \cdot\langle 2^+_k \parallel Q\parallel 0^+_1 \rangle \\
\label{eq:q5sum}
q_5 = & \sqrt{\frac{7}{10}} \  & |\sum_{i,j,k,l} 
		\langle 0^+_1 \parallel Q\parallel 2^+_i\rangle 
		\langle 2^+_i \parallel Q\parallel 2^+_j\rangle 
		\langle 2^+_j \parallel Q\parallel 0^+_k \rangle \nonumber \\
 & & 		\cdot\langle 0^+_k \parallel Q\parallel 2^+_l\rangle 
		\langle 2^+_l \parallel Q\parallel 0^+_1 \rangle| \\
\label{eq:q6sum}
q_6 = & \frac{7}{10} & \sum_{i,j,k,l,m} 
		\langle 0^+_1 \parallel Q\parallel 2^+_i\rangle 
		\langle 2^+_i \parallel Q\parallel 2^+_j\rangle 
		\langle 2^+_j \parallel Q\parallel 0^+_k \rangle \nonumber \\
 & & 		\cdot\langle 0^+_k \parallel Q\parallel 2^+_l\rangle 
		\langle 2^+_l \parallel Q\parallel 2^+_m\rangle
		\langle 2^+_m \parallel Q\parallel 0^+_1 \rangle \ ,
\end{eqnarray}

involving reduced matrix elements between $0^+$ and $2^+$ states only. In 
general only the lowest states contribute to the sums because convergence 
of the $Q$-phonon expansion of nuclear states is fast \cite{Pie98}. Matrix 
elements between nuclear states that differ by several $Q$-phonons are 
usually small \cite{Ots94}.

In the model of a quadrupole deformed rotor analytical expressions
for $E2$ matrix elements and thus for shape invariants can be obtained. In the 
rigid rotor the shape invariants are functions of the fixed deformation 
parameters $\beta$ and $\gamma$. If we assume a non-rigid deformation, we can
give expressions for the shape invariants as
\begin{eqnarray}
\label{eq:defbeteff}
q_2 & = & \left(\frac{3ZeR^2}{4\pi}\right)^2 \ \langle \beta^2 \rangle \equiv \left(\frac{3ZeR^2}{4\pi}\right)^2 \ {\beta_{\rm eff}}^2 \\
\label{eq:defgameff}
K_3 & = & \frac{\langle \beta^3 \cos{3\gamma} \rangle}
	  {\langle \beta^2 \rangle^{3/2}} \equiv \cos{3\gamma_{\rm eff}} \\ 
\label{eq:k4gambet}
K_4 & = & \frac{\langle \beta^4 \rangle}{\langle \beta^2 \rangle^2} \\
\label{eq:k5gambet}
K_5 & = & \frac{\langle \beta^5 \cos{3\gamma} \rangle}
	  {\langle \beta^2 \rangle^{5/2}}\\
\label{eq:k6gambet}
K_6 & = & \frac{\langle \beta^6 \cos^2{3\gamma} \rangle}
	  {\langle \beta^2 \rangle^3} \ ,
\end{eqnarray}
explicitly using expectation values of $\beta$ and $\cos{3\gamma}$. 
We can define effective values of the deformation 
parameters $\beta_{\rm eff}$ and $\gamma_{\rm eff}$ by 
Eqs. (\ref{eq:defbeteff},\ref{eq:defgameff}). 
The parameter 
$\gamma_{\rm eff}=\frac{1}{3}\arccos{K_3}$ is given in Table \ref{tab:klimits} 
for the appropriate dynamical symmetry 
limits of the IBM, where $SU(3)$ corresponds to a symmetric rigid rotor,
$O(6)$ to a $\gamma$-soft nucleus with maximal triaxiality and $U(5)$ to
a vibrator.

The shape invariants 
are measures of effective deformation parameters and their fluctuations. 
This is made more explicit by defining the following quantities as measures 
of the fluctuations of $\beta$ and $\cos{3\gamma}$:
\begin{eqnarray}
\label{eq:sigma4Def}
\sigma_{\beta} & = & \frac{\langle \beta^4 \rangle - \langle \beta^2 \rangle^2}
	       {\langle \beta^2 \rangle^2} = K_4 - 1 \\
\label{eq:sigma6Def}
\sigma_{\gamma} & = & \frac{\langle \beta^6 \cos^2{3\gamma} \rangle - 
	       \langle \beta^3 \cos{3\gamma} \rangle^2}
	       {\langle \beta^2 \rangle^3} = K_6 - {K_3}^2
\end{eqnarray}

\begin{figure}[hbt]
\epsfxsize 8.0cm
   \centerline{\epsfbox{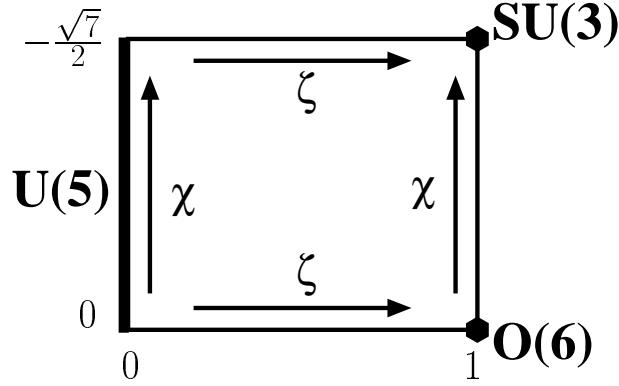}}
\caption{The ECQF-square for transition matrix elements in the IBM-1.
\protect$U(5)$ converges to one single point for \protect$N=\infty$.}
\label{fig:ECQFsquare} 
\end{figure}

Using expressions (\ref{eq:q2sum}--\ref{eq:q6sum},\ref{eq:sigma4Def},
\ref{eq:sigma6Def}) one can analytically calculate the shape invariants in
the dynamical symmetry limits of the IBM-1.
In this paper we will employ the Extended Consistent Q Formalism (ECQF)
\cite{Lip85} of the IBM-1, using the IBM-1 Hamiltonian
\begin{equation}
\label{eq:hecqf}
H_{ECQF} = a \ \left[(1-\zeta) \ n_d - \frac{\zeta}{4N} \ Q^{\chi} \cdot Q^{\chi}\right] \ ,
\end{equation}
with $Q^{\chi}$ taken from Eq. (\ref{eq:qchi}). This simple Hamiltonian
contains three parameters ($a$,$\zeta$,$\chi$). While one parameter
($a$) sets the absolute energy scale, the wave functions depend only on
two structural constants ($\zeta$,$\chi$). For a given nucleus 
the boson number $N$ is fixed. The ECQF-Hamiltonian
covers the three dynamical symmetry limits as indicated in Fig. 
\ref{fig:ECQFsquare}. We note that the structural parameter $\chi$ appearing 
in the shape invariants through the $E2$ transition operator is fixed in the 
$SU(3)$ limit ($\chi$=$-\sqrt{7}/2$) and in the $O(6)$ limit ($\chi$=$0$) 
while it is unspecified by the Hamiltonian in the $U(5)$ limit.
Therefore shape invariants in the $U(5)$ limit are functions of the
structural parameter $\chi$.
The analytical expressions for the shape invariants and the fluctuations
in the $U(5)$ limit as functions of the boson number $N$ and the structural
parameter $\chi$ are
\begin{eqnarray}
\label{eq:k3nchi}
K_3^{U(5)} & = & \sqrt{\frac{7}{10}} \ \frac{1}{\sqrt{N}} \ |\chi| \\
\label{eq:k4nchi}
K_4^{U(5)} & = & \frac{7}{5} \ \left( 1 - \frac{2}{7N}\right) \\
\label{eq:k5nchi}
K_5^{U(5)} & = & \sqrt{\frac{7}{2}} \ \frac{(11N - 6)}{(5N)^{3/2}} |\chi| \\
\label{eq:k6nchi}
K_6^{U(5)} & = & \frac{21}{25} \ \frac{(N-1)}{N^2} \ 
	        \left(3\chi^2+N-2+\frac{5N\chi^2}{6(N-1)}\right) \\
\label{eq:sigbnchi}
\sigma_{\beta}^{U(5)} & = & \frac{2}{5} \ \left( 1 - \frac{1}{N}\right) \\
\label{eq:siggnchi}
\sigma_{\gamma}^{U(5)} & = & \frac{21}{25} \ 
		\frac{(N-1)}{N^2} \ \left(3\chi^2+N-2\right) \ .
\end{eqnarray}
For completeness we also give $K_6$ as a function of the boson number $N$
in the $O(6)$ limit
\begin{equation}
\label{eq:k6o6n}
K_6^{O(6)} = \frac{1}{3} \ \frac{(N-2)(N-1)(N+5)(N+6)}{[N(N+4)]^2} \ .
\end{equation}
These results enlarge the well known symmetry triangle for wave functions
of the IBM-1 to a structural ECQF-square for shape invariants and thus
for the interpretation of nuclear shapes. This fact is illustrated in 
Fig. \ref{fig:ECQFsquare}. The use of a similar rectangular
representation of the parameter space has also been suggested by
Bucurescu {\em et al.} \cite{Buc99}. Known typical examples for particular
points of the ECQF-square are, e.g., $^{172}$Yb for 
$SU(3)$, $^{196}$Pt for $O(6)$, $^{116}$Cd for $U(5)$ with $\chi$=$0$
and $^{152}$Sm for large $\chi$ and moderate values of $\zeta$ 
(see \cite{Cas98,Zam99,Iac98} and discussion below).

From Eqs. (\ref{eq:k3nchi},\ref{eq:k5nchi},\ref{eq:k6nchi},\ref{eq:siggnchi})
we note that the necessary extension of the IBM-1 symmetry triangle
to the ECQF-square is a finite-N-effect, because the shape invariants
of $U(5)$ wave functions converge in the limit $N\rightarrow\infty$
for any value of $\chi$. Table \ref{tab:klimits}
shows the values of the shape invariants and their fluctuations in the
dynamical symmetry limits of the IBM-1 for an infinite boson number
$N=\infty$. Only the quantities given in Eqs. 
(\ref{eq:k3nchi}--\ref{eq:k6o6n}) depend on the boson number $N$.

As we would expect the values of $\gamma_{\rm eff}$ in the
symmetry limits are $0^{\circ}$ and $30^{\circ}$ while the effective
triaxiality fluctuates in the $O(6)$ and the $U(5)$ limits.
In the $SU(3)$ rigid rotor and $O(6)$ $\gamma$-soft limits the $\beta$ 
deformation is rigid while it fluctuates in the $U(5)$ vibrator limit.

We have discussed the shape invariants and 
fluctuations in the dynamical symmetry limits of the IBM using 
analytical expressions. 
These values provide useful benchmarks for the geometrical interpretation of
IBM ground state wave functions. However, the dynamical symmetry limits of 
the IBM and the corresponding geometrical models are idealised, analytically
solvable limits. More accurate descriptions of the low energy structure of
collective nuclei can usually be obtained by IBM Hamiltonians outside the
dynamical symmetry limits. To gain insight in the structure of
actual nuclei the quantities of interest have been calculated 
between the symmetry limits, using the ECQF-Hamiltonian (\ref{eq:hecqf}). 
The shape invariants and their fluctuations have been 
calculated gridwise over the whole IBM parameter space for $N=10$ bosons
as functions of the structural parameters $\zeta$ and $\chi$.
All calculations have been performed by diagonalizing the Hamiltonian 
numerically using the computer code PHINT \cite{phint}. Calculations
of the shape invariants have been done by a FORTRAN code (QINVAR)
which evaluates the PHINT output.

\begin{figure}[hbt]
\epsfxsize 17.0cm
   \centerline{\epsfbox{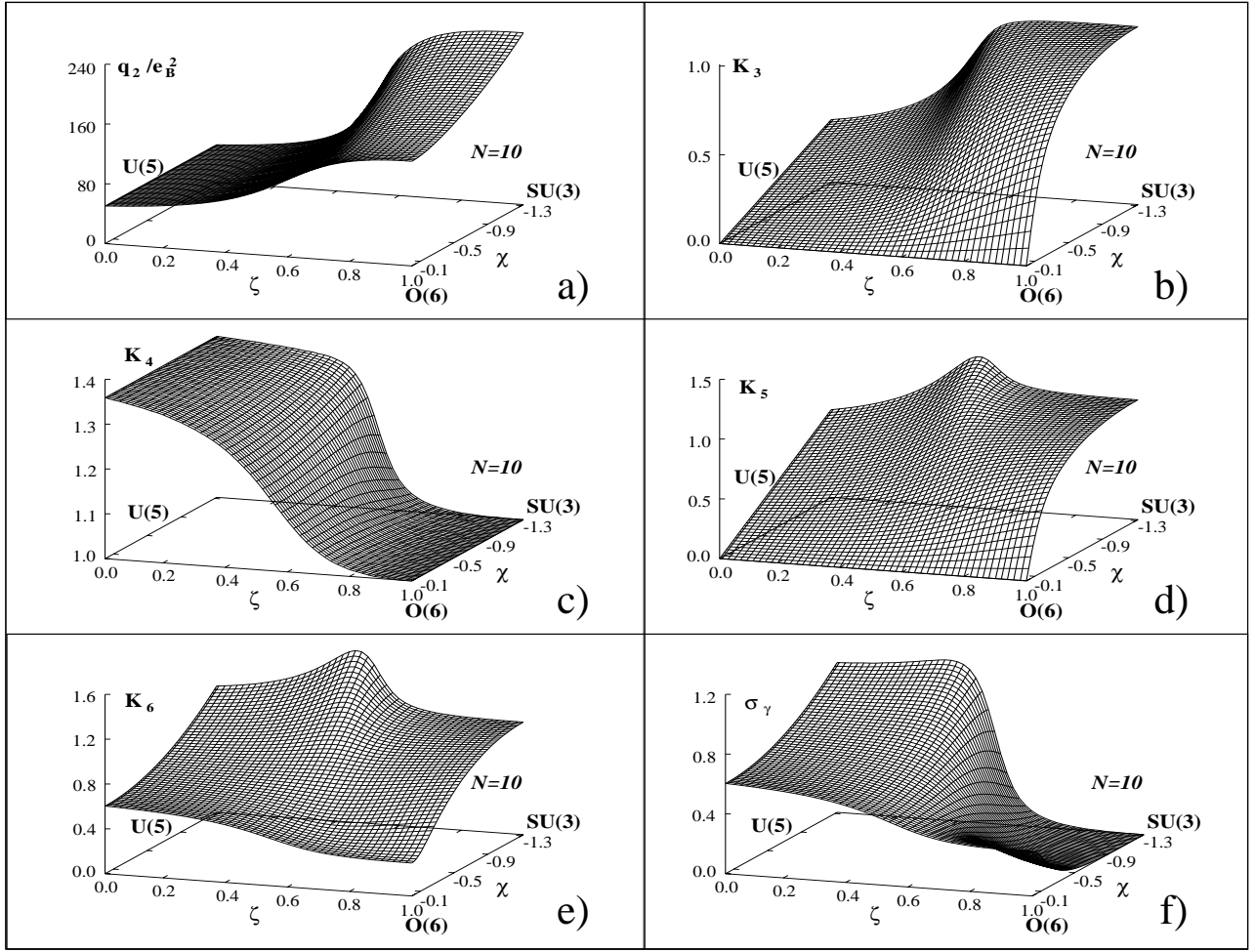}}
\caption{a)-e): Shape invariants $q_2$ and $K_3$ - $K_6$ c): The invariant 
	 \protect$K_4$ also determines the 
	 fluctuation \protect$\sigma_{\beta}=K_4-1$. f): The fluctuation 
	 \protect$\sigma_{\gamma}$.
	 All quantities shown in a)-f) are calculated gridwise over the 
	 ECQF parameter space of the IBM for \protect$N=10$ bosons.}
\label{fig:kgrids} 
\end{figure}

All quantities behave smoothly and one obtains an impression of
how the quantities vary outside of the dynamical symmetry limits. Fig. 
\ref{fig:kgrids} represents the numerical results of this work presenting
the variation of the most important quadrupole invariants over
all ranges of structure. The behaviour of the invariants, $q_2$, $K_3$-$K_6$,
between the symmetries is interesting. Strong variations towards
and for deformed nuclei are typical. The invariant $K_4$, which is related to
fluctuations in $\beta$ via Eq. (\ref{eq:sigma4Def}), is one of the few
observables that can distinguish $U(5)$ from $O(6)$. This can be useful in
newly accessible exotic nuclei since $K_4$ can be approximately obtained
from the simple expression \cite{Jol97}
\begin{equation}
\label{eq:k4approx}
K_4 \approx \frac{7}{10} \frac{B(E2;4_1^+ \rightarrow 2_1^+)}{B(E2;2_1^+ \rightarrow 0_1^+)} \equiv K_4^{\rm appr.} \ ,
\end{equation}
which involves two observables, easily measured, e.g., by Coulomb
excitation experiments. The approximation (\ref{eq:k4approx}) is valid
within about $10\%$ for the ECQF-square and boson numbers $N\ge 5$.
This was numerically checked for the whole ECQF-square and for boson
numbers $N=5,7,10,16$.
For a detailed analysis an experimental value of $K_4^{\rm appr.}$ can
serve as a benchmark for starting points of numerical IBM calculations,
which can be optimized to reproduce the measured transition strengths.
The actual value of $K_4$ can then be determined from the complete set of 
calculated  $E2$ transition matrix elements.

For large $N$, $K_6$ and $\sigma_{\gamma}$ are quite different in $U(5)$ 
and $O(6)$ which is evident from Table \ref{tab:klimits}. 
The bottom right panel of Fig. \ref{fig:kgrids} shows $\sigma_{\gamma}$,
which gives the fluctuations in $\gamma$, gridwise over the full 
structural range. Note, however, that Fig. \ref{fig:kgrids} is calculated 
for a finite boson number which lowers the value of $K_6$ and 
$\sigma_{\gamma}$ in the $U(5)$ and $O(6)$ limits as seen from 
Eqs. (\ref{eq:k6nchi},\ref{eq:siggnchi},\ref{eq:k6o6n}).
In the $SU(3)$ limit $\sigma_{\gamma}$ vanishes, which characterizes 
the $SU(3)$ limit as a model for a rigid rotor, also in the $\gamma$ 
degree of freedom. In contrast non-vanishing triaxiality 
fluctuations occur in the $U(5)$ and $O(6)$ limits, indicating that these 
limits and the whole transitional region between them model $\gamma$-soft 
nuclei. 

Finally, we note that all shape invariants, especially $\sigma_{\gamma}$,
change strongly between $SU(3)$ and $U(5)$-like values
in an unusual region of the IBM-1 parameter space, namely for moderate values 
of $\zeta$ and $\chi$=$-\sqrt{7}/2$. Interestingly,
this is just the region appropriate to the nucleus $^{152}$Sm 
($\zeta$=$0.57$,$\chi$=$-\sqrt{7}/2$) \cite{Cas98,Zam99}. The case of 
$^{152}$Sm  is currently under active discussion and it seems that it shows a 
certain degree of shape coexistence between spherical and deformed shapes 
with large effective triaxiality \cite{Iac98}.

Above, we discussed the numerical calculation of  the exact shape invariants
within the $sd$-IBM-1 parameter space, using the ECQF-Hamiltonian 
(\ref{eq:hecqf}). One aspect of this work is to establish the shape invariants
as a convenient link between the geometrical model and any other nuclear
structure model which is able to calculate $E2$ transition matrix elements.
Here we have chosen the algebraical IBM. Our ansatz is alternative to the
intrinsic state formalism by Ginocchio and Kirson \cite{GiKi80} which was
used much earlier to link the IBM Hamiltonian to the geometrical Bohr
Hamiltonian.

We note that the effective values of the shape parameters $\beta_{\rm eff}$
and $\gamma_{\rm eff}$ do in general not exactly coincide with the minima of
a corresponding energy surface for the ground state in the deformation 
parameter plane. However, the shape invariants can easily be used to 
compare the predictions from different nuclear models in a geometrically 
transparent way.

In principle, the shape invariants can also be measured directly 
from extensive nuclear structure data, providing a direct test of nuclear
structure models. Much more intriguing is the common case
when only a few key observables, like $E2$ branching ratios from 
low-lying $0^+$ states and $2^+$ states, are known experimentally 
and when a phenomenological nuclear structure model, like the IBM, 
can be used to extrapolate the data to a complete set of $E2$ 
transition matrix elements.

To summarize, we have presented analytic expressions for moments up to sixth 
order of the quadrupole operator in the ground state and we have given 
definitions for the lowest shape invariants up to $K_6$. The shape invariants 
were calculated analytically in the dynamical symmetry limits of the IBM-1.
Formulae were 
given to derive effective deformation parameters and their fluctuations from 
shape invariants, and thus from IBM-1 calculations. A study, using the 
ECQF-Hamiltonian (\ref{eq:hecqf}), of the behaviour of the shape invariants 
over a full range of structures has been performed for the first time. It 
shows the smooth but yet widely varying behaviour of the invariants. Thus they
can be used to determine the properties of nuclei by comparing the calculated 
invariants to experimentally obtained values or to results of fits. Moreover,
approximate values of these invariants can be obtained experimentally
simply from $B(E2)$ values involving just the $2_1^+$, $2_2^+$ and $4_1^+$ 
states, and, for $K_5$ and $K_6$, a $B(E2)$ branching ratio from the
appropriate excited $0^+$ states.

The invariant $K_4$, as well as the fluctuation $\sigma_{\gamma}=K_6-K_3^2$, 
are of special interest as they allow to distinguish between $O(6)$ and $U(5)$ 
symmetries which can be difficult otherwise \cite{Lev86}. Finally, the 
values of $\sigma_{\gamma}$ change most rapidly
for IBM-1 Hamiltonians that show shape coexistence.

For fruitful discussions the authors thank A. Gelberg, T. Otsuka and
N.V. Zamfir. This work has been partly supported by
the Deutsche Forschungsgemeinschaft under Contract Nos. Br 799/9-1 and 
Pi 393/1-1, and by the U.S. DOE under Grant No. DE-FG02-91ER40609.

\begin{table}[htb]
\caption{Shape invariants and fluctuations, calculated for $N=\infty$ 
in the $sd$-IBM-1 for the dynamical symmetry limits using analytical 
expressions. For comparison corresponding values of an effective $\gamma$ 
deformation are given.}
\begin{tabular}{c|ccc}
$N=\infty$ & U(5) & SU(3) & O(6) \\
\\
\hline
$\gamma_{\rm eff}$ & $30^{\circ}$ & $0^{\circ}$ & $30^{\circ}$ \\
\\
$K_3$ & 0 & 1 & 0 \\
$K_4$ & $\frac{7}{5}$ & 1 & 1 \\
$K_5$ & 0 & 1 & 0 \\
$K_6$ & $\frac{21}{25}$ & 1 & $\frac{1}{3}$ \\
\\
$\sigma_{\beta}$ & $\frac{2}{5}$ & 0 & 0 \\
$\sigma_{\gamma}$ & $\frac{21}{25}$ & 0 & $\frac{1}{3}$ \\
\end{tabular}
\label{tab:klimits}
\end{table}

\end{document}